# All-fiber polarization-manageable directional couplers

Liang Fang and Jian Wang*

*Wuhan National Laboratory for Optoelectronics, School of Optical and Electronic Information, Huazhong University of Science and Technology, Wuhan 430074, Hubei, China.*
*\* jwang@hust.edu.cn*

**Abstract:** Flexible management of light polarization is necessary to develop the technologies of optical fiber communications. However, it is still a great challenge to manipulate optical polarization in all-fiber-based devices. Here we present a novel structural design of three-core fiber couplers to create arbitrary polarization management by embedding a vector mode segregated core within the conventional fiber couplers. We take advantages of these coupling structures to achieve polarization split, rotation, and exchange, as three essential polarization managements for fiber-guided modes. We study these couplers based on coupled-mode theory and numerically simulations. The obtained results show that they manifest favorable operation performance on management of the fiber modal polarization states. Compared to previous couplers used for management of power, wavelengths and modal forms, this kind of fiber couplers introduces a new manageable degree of freedom for modal polarization. It is expected that the presented all-fiber polarization-manageable directional couplers might find potential applications in fiber optical communications and sensing systems.

**Introduction**

Fiber directional couplers have been widely used for power distribution, data uploading/downloading and exchange in fiber communications networks due to their high-efficiency coupling property, large working bandwidth and low manufacturing costs.[1] When employing controllable long-period gratings, the fiber couplers can implement tunable wavelength management for wavelength-division multiplexing (WDM) systems.[2,3] Additionally, in few-mode fiber (FMF) coupling systems, the fiber couplers can be exploited to selectively add/dropp higher-order linear polarization (LP) or orbital angular momentum (OAM) modes, and thus potentially applied to mode-division multiplexing (MDM) systems to increase the ever-increasing bandwidth demand of fiber-based communications.[3-9] Furthermore, two orthogonal polarization states of fiber-guided modes, acting as another fundamental degree of freedom of light, are usually employed for polarization-division multiplexing (PDM), which is combined with WDM and MDM to further increase the channel capacity.[11,12] Thereby, it is a laudable goal to directly adopt the fiber couplers to selectively couple different modal polarizations. As summarized in Figs. 1(a)-(c), the modified fiber couplers have been previously used to control optical power, wavelength, and modes. However, so far there have been limited research efforts of all-fiber polarization-manageable couplers to selectively achieve desired modal polarizations.

As well known, as for optical modes in fibers (i.e. circularly symmetrical waveguides), unlike the stripe silica-based waveguides where TM and TE modes can be easily separated, the fiber-guided odd and even HE/EH vector mode components with orthogonal polarization states tend to be degenerated because of nearly the same effective refractive indices between them, so does the first vector mode group, including the special cylindrical vector modes $TM_{01}$, $TE_{01}$, and even and odd $HE_{21}$ that may degenerate into $LP_{11}$ modes. It gives rise to the polarization-insensitivity of the conventional fiber couplers that usually couples the fundamental mode $HE_{11}$ as the most widely used mode in existing fiber communications systems. As a result, it limits their applications to modal polarization manipulation in fiber-based devices, such as polarization split and rotation, polarization-selectively coupling in a PDM system, etc. As for polarization beam splitters (PBS) as one of essential components in photonic integrated circuits and optical communications,[13,14] various types of PBS design have been proposed based on $LiNbO_3$,[15] semiconductors,[16] two dimensional grating coupler,[17] asymmetrical silicon-based directional coupler,[18,19] birefringent-fiber coupler,[20] photonic crystals,[21,22] etc. The birefringent-fiber-based and photonic-crystals-based PBSs belong to all-fiber polarization-manageable devices, nonetheless, they have poor compatibility with the existing fiber-based communication networks. The commercially available fiber polarization splitter is inlaid with a tiny polarization beam splitting prism, so strictly speaking it is not an all-fiber device. Therefore, it is favorable to design a simple and feasible all-fiber-based devices that can achieve flexible polarization-selective modal manipulation, and are fully compatible with the existing fiber communications systems as well.

In this paper, we propose a novel three-core coupling structural design to realize arbitrarily polarization-selective modal coupling. A high-contrast index ring core is embedded between two cores in the conventional couplers. This ring core can support several separated vector modes, including the exploitable cylindrical vector modes $TM_{01}$, $TE_{01}$, and even and odd $HE_{21}$, and significantly can lift the polarization mode degeneracy. It is well known that this first vector mode group in FMFs can be easily divided into the separated vector mode components in terms of the effective refractive index difference by increasing refractive index contrast between core and surroundings.[23-25] The radially polarized $TM_{01}$ and azimuthally polarized $TE_{01}$ modes have some distinct features and can supply many potential applications.[26] Here if the $TM_{01}$ or $TE_{01}$ mode supported in the ring core is matched with the mode coupling within the conventional fiber couplers, it will allow coupling between specific polarized modes, and block coupling for its orthogonal counterparts. What is more, the matched vector mode in the ring core can change the polarization orientation of coupled $HE_{11}$ mode from one SMF to another SMF if these two SMFs are placed at two different relative positions around this ring core. The polarization-selective modal coupling structures designed in this paper are based on these significant coupling properties. We investigate the coupling mechanism of these three-core coupling structures based on coupled-mode theory, and discuss in detail the polarization-dependent coupling feature, as well as coupling crosstalk between the single-mode core and the ring core. Three essential polarization management types for fiber-guided modes, i.e. modal polarization split, rotation, and exchange are successfully implemented using these coupling structures. We numerically simulate these devices and investigate the conversion efficiency, work bandwidth, and the coupled polarization purity in detail. These three-core structures of all-fiber polarization manageable couplers can be fabricated by means of side-polishing single fibers and then gluing them together.[27,28] We believe that they may introduce a new manageable degree of freedom for modal polarization as shown in Fig. 1(d), relative to the conventional fiber couplers used for management of optical power, wavelengths and modal forms. It is expected that these new kinds of directional couplers might find wide applications in fiber communications and sensing systems, etc.

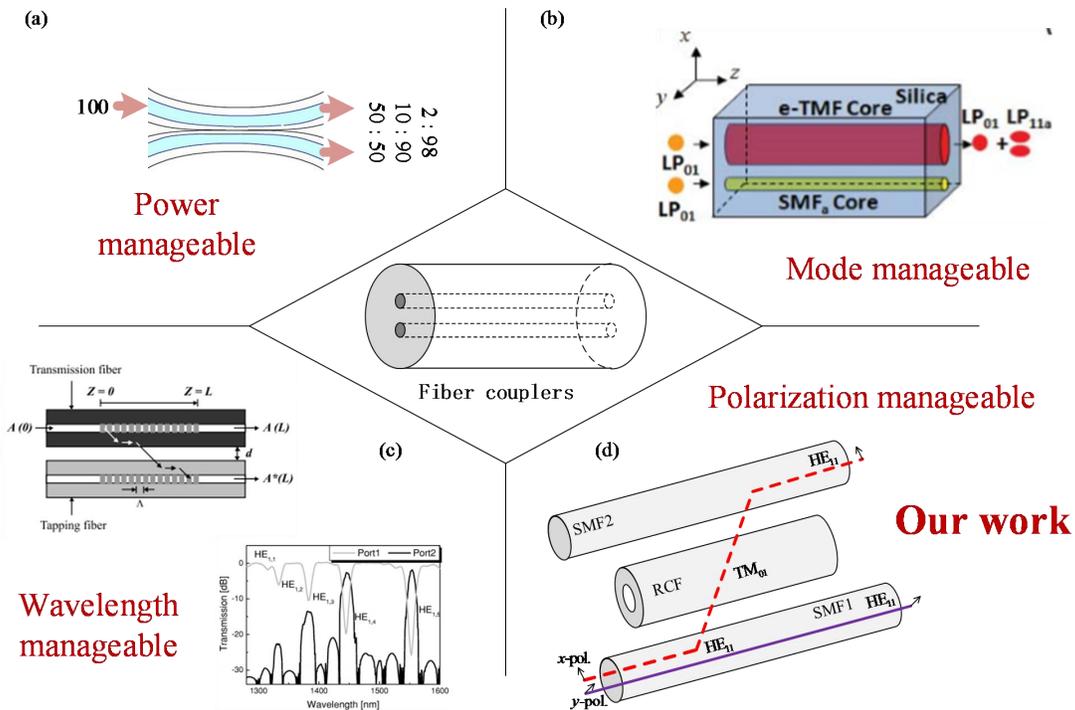

Fig. 1 Modified all-fiber directional couplers for multi-dimensional management of light with respect to (a) modal power, (b) wavelengths, (c) modal forms, and (d) polarizations.

## Materials and Methods
**Coupled-mode theory about transitive coupling among three modes**

Coupled-mode theory is an available way to obtain quantitative information about power transfer in any parallel fiber systems. The solution of a two-fiber system is a commonplace, especially for two identical single-mode fiber (SMF). If mode coupling occurs among three or more modes in a more complex system, the coupled-mode equations will be complicated so that it is difficult to get analytical solutions, thereby we present a convenient numerical method to treat with it in supplementary.

For polarization-dependent mode coupling in the proposed fiber system as shown in Fig. 2(a), we theorized the case of three-mode transitive coupling in a three-core fiber system and get an analytical solution. The evolution of the electrical fields in three-coupled mode systems have been investigated, [29,-31] and can be described by the coupled-mode equations

$$\frac{dA_1}{dz} + j\beta_1 A_1 = j\kappa_{12} A_2 , \qquad (1)$$

$$\frac{dA_2}{dz} + j\beta_2 A_2 = j\kappa_{21} A_1 + j\kappa_{23} A_3 , \qquad (2)$$

$$\frac{dA_3}{dz} + j\beta_3 A_3 = j\kappa_{32} A_2 , \qquad (3)$$

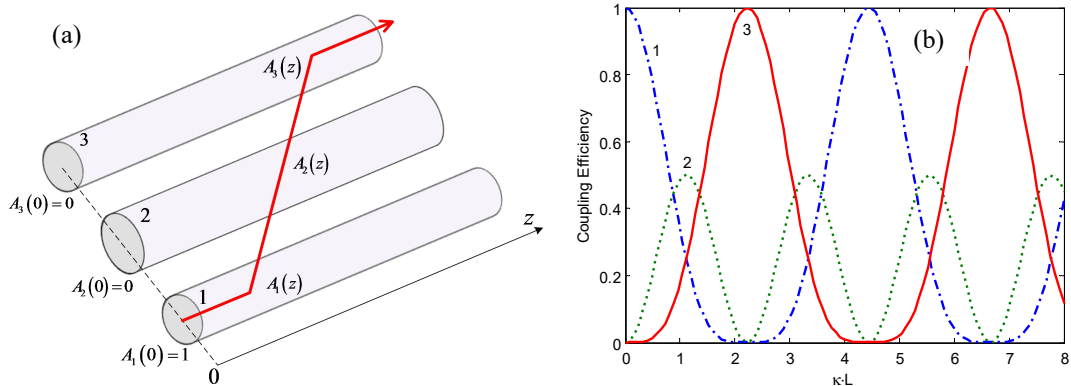

Fig. 2 (a) Sketch of mode transitive coupling among three modes in a three-core fiber structure. (b) Power transfer among three modes versus the product of coupling coefficient and coupling length.

where $A_i$ denotes the complex amplitude of electric field of each mode along the $+z$ direction, $\beta_i$ indicates the propagation constant of corresponding modes and defined by $\beta_i = 2\pi n_{eff}^i / \lambda$ with $\lambda$ being the wavelength and $n_{eff}^i$ being the effective refractive index ( $i = 1, 2, 3$ ). $\kappa_{12} \simeq \kappa_{21}$ represents the coupling coefficient between mode1 in core1 and mode2 in core2, and $\kappa_{23} \simeq \kappa_{32}$ between mode2 and mode3, respectively. The coupling coefficient has a remarkable property of polarization-dependence, and it is a key to exploit the polarization-management devices proposed in this paper. The polarization-dependence and reciprocal relationship of these coupling coefficients are discussed in detail in supplementary. In fact, mode direct coupling between mode1 in core1 and mode3 in core3 may be produced, but in contrast with adjacent coupling between mode1 and mode2 or mode2 and mode3, the coupling strength is so weak that it can be neglected. This is because that the coupling coefficient between them is $\kappa_{13} \ll \kappa_{12}, \kappa_{23}$ due to the too large distance between core1 and core3. Assuming that fiber cores 1 and 3 have identical sizes, $\beta_1 = \beta_3$, under the boundary conditions of $A_1(0) = 1$, $A_2(0) = 0$, $A_3(0) = 0$, and $dA_3(0)/dz = 0$, one can get the solutions to Eqs. (1) to (3) as follows

$$A_1(z) = \frac{\kappa_{23}^2}{\chi^2} + \frac{\kappa_{12}^2}{\chi^2} \cos\gamma z \cdot e^{j\delta z} - \frac{j \cdot \kappa_{12}^2 \cdot \delta}{\gamma \cdot \chi^2} \sin\gamma z \cdot e^{j\delta z} , \qquad (4)$$

$$A_2(z) = j\frac{\kappa_{12}}{\gamma}\sin\gamma z \cdot e^{-j\delta z}, \tag{5}$$

$$A_3(z) = -\frac{\kappa_{12}\cdot\kappa_{23}}{\chi^2} + \frac{\kappa_{12}\cdot\kappa_{23}}{\chi^2}\cos\gamma z \cdot e^{j\delta z} - \frac{j\kappa_{12}\cdot\kappa_{23}\cdot\delta}{\gamma\cdot\chi^2}\sin\gamma z \cdot e^{j\delta z}, \tag{6}$$

where $\delta = (\beta_1 - \beta_2)/2$ being the resonance factor, $\chi = \sqrt{\kappa_{12}^2 + \kappa_{23}^2}$, and $\gamma = \sqrt{\delta^2 + \chi^2}$. It meets the condition of total energy conservation of $|A_1(z)|^2 + |A_2(z)|^2 + |A_3(z)|^2 = 1$. If fiber cores 1 and 3 are symmetric with respect to the middle core2, two coupling coefficients will be equal, i.e. $\kappa_{12} = \kappa_{13} = \kappa$, under the condition of full resonance, i.e. $\delta = 0$, the power evolution of each coupled mode along the $+z$ direction can be obtained as

$$P_1(z) = |A_1(z)|^2 = \cos^4\left(\frac{\sqrt{2}}{2}\kappa z\right), \tag{7}$$

$$P_2(z) = |A_2(z)|^2 = \frac{1}{2}\cdot\sin^2\left(\sqrt{2}\kappa z\right), \tag{8}$$

$$P_3(z) = |A_3(z)|^2 = \sin^4\left(\frac{\sqrt{2}}{2}\kappa z\right). \tag{9}$$

The full coupling efficiency from core1 to core3 through core2 occurs under $\kappa\cdot z = \sqrt{2}(1+2n)\pi/2$, whereas the maximum modal power for the middle mode in core 2 can be obtained in the case of $\kappa\cdot z = \sqrt{2}(1+2n)\pi/4$, ($n = 0,1,2...$). This is a significant coupling rule for the mode coupling among three symmetrical cores and it is very different from the coupling case between two modes in conventional fiber couplers where the full coupling efficiency is produced when $\kappa\cdot z = (2n+1)\pi/2$. The power transfer among these three modes are calculated as a function of the product of coupling coefficient and coupling length, as shown in Fig. 2(b). It clearly shows that modal power in one fiber can be transferred into another fiber through the middle fiber, and the transfer efficiency approaches 100% in theory. The three-coupled mode systems have been studied on adiabatic elimination-based coupling control in densely packed subwavelength waveguides,[32, 33] where the coupling process distinctly follows the principle revealed here.

**Structural design and polarization-management principle.**

In the presented design, the hollow high-contrast index ring core serves as the middle core in the three-core coupling system. Two cores on each side correspond to the normal SMFs with core refractive index $n_0 = 1.448$. Three cores are packed with a mutual cladding of SiO$_2$ materal, The ring core has a high step index of $\Delta \simeq 2.3\%$, corresponding to $n_2 = 1.478$, and the inner layer is air, i.e., $n_1 = 1$. Previous works have reported the fabrication of this high-contrast index ring core fiber.[9,34,35] Firstly, we present the variation of effective refractive indices of the fundamental mode HE$_{11}$ in independent SMF and several high-order vector modes in independent RCF as a function of core radius at the wavelength of 1550nm as shown in Fig. 3A. We get the effective refractive indices map by numerically solving the vector eigenmode equations of two and three layers of waveguide models that correspond to SMF and RCF, respectively.[36,37] The size design and refractive indices distribution of SMF and core are shown in the insets. The inner and outer radii of RCF are conditioned to $a_2 = 1.5a_1$, and we vary the inner radius $a_1$ to make the indices equal between HE$_{11}$ in SMF and desired vector modes in RCF. One can see from Fig. 3A that the effective refractive indices of the supported first (TM$_{01}$, TE$_{01}$ and HE$_{21}$) and second (HE$_{31}$ and EH$_{11}$) vector mode groups are completely segregated in the high-contrast index ring core. The segregation degree of the first one is higher than the second one, and the effective refractive indices difference $\Delta n_{eff}$ between any two of them is more than $5\times10^{-3}$. Under the condition of fixed core radius $a_0 = 4.6\mu m$ for SMF, if one wants to get coupled TM$_{01}$ mode in RCF, the inner radius of RCF needs to be fabricated as or tapered into the marked value $a_1$ in Fig. 3A to satisfy the resonance condition of vector mode coupling. If getting $HE_{21}$ mode, it corresponds to the marked value $a_1'$.

The vector mode coupling highly depends upon the modal polarization states. We exhibit the cases of several polarization-dependent excitation from $HE_{11}$ mode in SMF to vector modes in RCF in Fig. 3**B**. The x-polarized $HE_{11}$ mode can be coupled into $TM_{01}$ and odd HE/EH modes, whereas the y-polarized $HE_{11}$ mode into $TE_{01}$ and even HE/EH modes. The strong coupling occurs in the case of uniform polarization of electric fields for the two coupled modes in the overlap regions, which is determined by the definition of coupling coefficients based the coupled-mode theory. The detailed discussion is given in the supplementary. Especially, when the input $HE_{11}$ mode has a circular polarization state, the excited HE/EH modes manifest OAM carrying modes. The polarization handness of OAM modes is consistent with input $HE_{11}$ mode, but the spatial phase is characterized by helical spatial distribution, different from the planar phase of input $HE_{11}$.[9] This polarization-dependent coupling feature for vector mode excitation is analogous to azimuthal-dependent coupling for higher-order LP modes in FMF coupling systems that can be used to selectively excite odd and even LP modes, but cannot differentiate two orthogonal polarization states for each odd or even LP mode.[4,5]

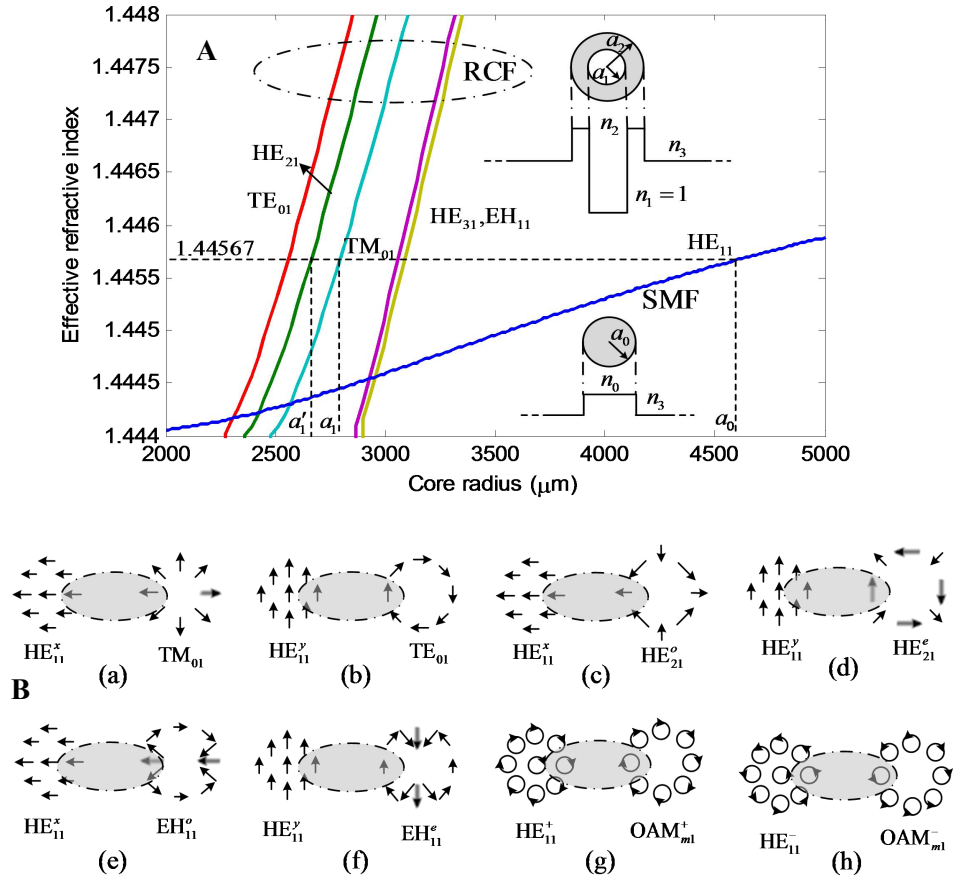

Fig. 3 **A** The effective refractive indices map of the fundamental mode $HE_{11}$ in SMF and several high-order vector modes in RCF versus core radii at the wavelength of 1550nm. **B** Sketches of polarization-dependent coupling where the gray ellipses show the polarization uniform at the close sides of two coupled modes.

Based on the principle of reciprocal coupling, mode coupling from vector modes in RCF to $HE_{11}$ mode in SMF is also polarization-dependent. In our proposed three-core coupling systems, if two single-mode cores are placed at two different relative positions around the ring core, but with the same distance away from this ring core, the $HE_{11}$ mode coupling from one single-mode core to another single-mode core through the ring core will change the polarization orientation. As for different middle vector mode matched cases, we present different evolution cases of polarization orientation in Fig. 4A. It should be stressed that the TM$_{01}$ and TE$_{01}$ middle matched cases in Fig. 4A(a) and (b) can just work for the corresponding linear polarization input of HE$_{11}$, but does not work for its orthogonal counterpart, because the large effective index difference between TM$_{01}$ and TE$_{01}$ modes undoubtedly gives rise to the deturning between one of them and HE$_{11}$. In contrast, the HE$_{21}$ middle matched case can work for both two orthogonal polarization inputs of HE$_{11}$ because of nearly the same effective refractive indices of odd and even HE$_{21}$ modes in the ring core. As shown in Fig. 4A (c), we only present the odd HE$_{21}$ matched case, and it is analogous for the even case.

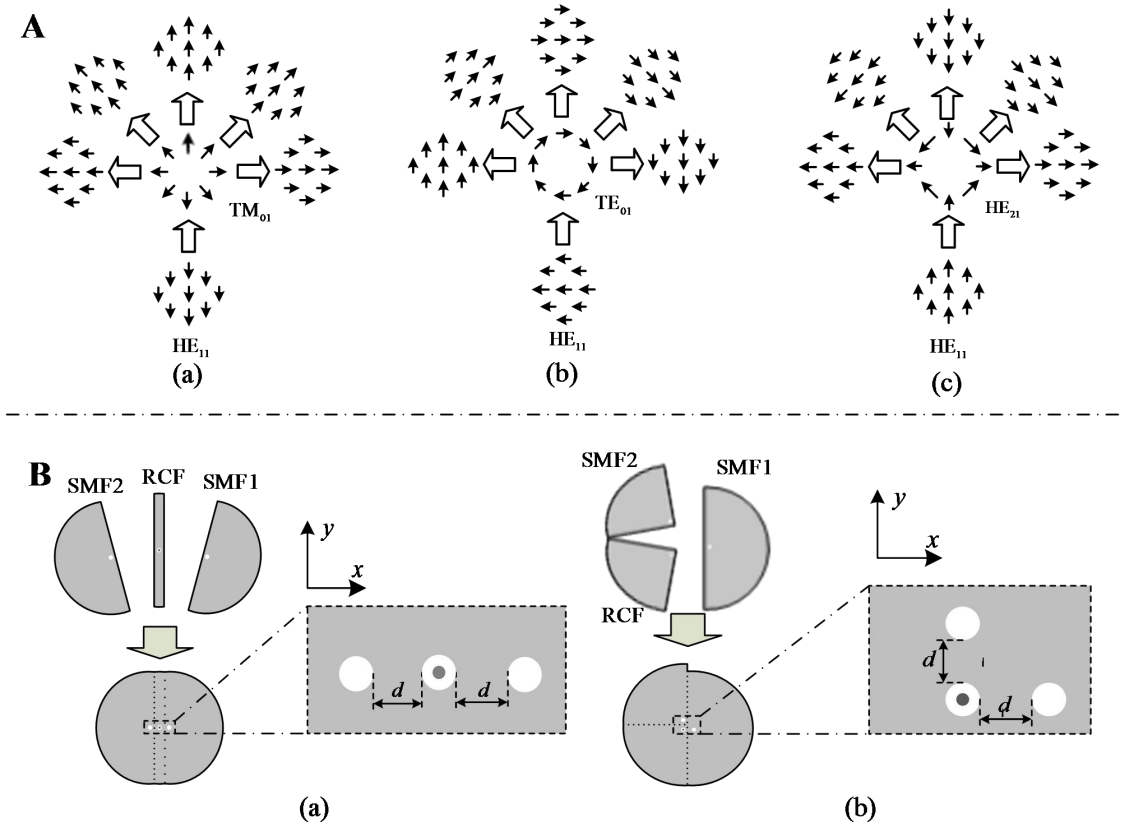

Fig. 4 **A** Three types of vector mode middle matched cases for polarization orientation evolution. (a) TM$_{01}$ matched case, (b) TE$_{01}$ matched case, (a) odd HE$_{21}$ matched case. **B** The feasible fabrication process and three-core fiber structures of all-fiber-based (a) polarization splitter and (b) polarization rotator or exchanger.

In this paper, based on the polarization-manageable principle of three-core coupling structures discussed above, we design three types of all-fiber polarization manageable couplers to correspondingly obtain three essential polarization manipulation functions, such as polarization split, rotation, and exchange. The layout of three cores for polarization splitting design is in a line, and TM$_{01}$ or TE$_{01}$ mode can be served as the middle matched mode in the ring core, whereas the polarization rotation and exchange couplers are designed as vertical layout of three cores. Either TM$_{01}$, TE$_{01}$, or HE$_{21}$ mode can be selected as the middle matched mode for polarization rotation and exchange. The main difference is that the TM$_{01}$ and TE$_{01}$ matched polariation rotation and exchange only work for a specific polarization input, but the HE$_{21}$ matched rotation and exchange can work for arbitrary polarization inputs. For the fabrication of these three-core fiber couplers, it seems to be difficult to adopt the methods of conventional fusing and tapering fibers. Here, we provide a feasible way to fabricate these kinds of fiber couplers. One can design them by bonding three side-polished independent SMFs and RCF with refractive index matching liquid or glue that is close to the fiber cladding.[27,28] Shown in Figs. 4B (a) and (b) are the possible fabrication process of all-fiber polarization splitter and polarization rotator or exchanger, respectively.

## Results and Discussion

### All-fiber polarization splitter

The core layout of all-fiber polarization splitting structure is sketched as shown in Fig. 5A(a), in which three cores are in a line, and four ports serve as inputs and outputs. Here we select the $TM_{01}$ mode as the middle matched mode in the ring core. Actually, the $TE_{01}$ mode can be also utilized. For the $TM_{01}$ matched mode, the splitted mode is x-polarized $HE_{11}$, whereas for the $TE_{01}$ matched mode, it turns to the y-polarized $HE_{11}$. The section view corresponds to Fig. 4B(a), where the distances between the ring core and two single-mode cores need to keep the same in order to get the symmetric coupling from SMF1 to SMF2 across RCF. Otherwise, the power may not be completely coupled from SMF1 to SMF2, resulted from the unbalanced coupling efficiency astride the RCF. When the $HE_{11}$ mode with arbitrary polarization inputs from the port1 in SMF1, the x-polarized component can be coupled to RCF and then to SMF2, and outputs from the port3 in SMF2, whereas the y-polarized component remains impervious, and outputs from the port2 in SMF1. We numerically simulate the polarization splitting of this coupling structure by finite-difference time-domain method. The fiber parameters of three cores set here are consistent with the dispersion map shown in Fig. 3A, but note that the inner radius of RCF is set as $a_1 = 2794$ nm at the working wavelength of 1550nm to satisfy the resonance condition. The distance between the ring core and each single-mode core is taken as $d = 8$ μm, and the required coupling length is about 1 cm. The results of power evolution map are shown in Fig. 5B, where the incident x-polarized $HE_{11}$ from the port1 in SMF1 can be completely coupled to SMF2 and output from the port3, as shown in Fig. 5B(a), in contrast, the y-polarized $HE_{11}$ propagation is nearly unaffected by this structure, as shown in Fig. 5B(b).

We then study the coupling properties of these coupling structures in terms of working bandwidth, coupling crosstalk and central wavelength shift due to fabrication errors. Firstly, we compare two kinds of calculated results between the analytical model and simulation model over a large bandwidth. In the analytical model, there are only three coupled modes, i.e. the incident x-polarized $HE_{11}$ from SMF1, middle mode $TM_{01}$ excited in RCF, and the targeted x-polarized $HE_{11}$ coupled in SMF2, shown as the thick lines in Fig. 5A(b). The coupling among these three modes follows the analytical expressions from Eqs. (4) to (5). In the simulation model, two additional coupling cases are also considered, including the coupling from incident x-polarized $HE_{11}$ from SMF1 to odd $HE_{21}$ mode in RCF and then to the x-polarized $HE_{11}$ coupled in SMF2, and the direct coupling from x-polarized $HE_{11}$ from SMF1 to x-polarized $HE_{11}$ in SMF2, indicated with the thin lines in Fig. 5A(b). The transferred energy of each coupled mode is monitored at the end of three cores when propagating along the full coupling length. We show two types of results in Fig. 5C(a), where the pure and dot-marked solid lines correspondingly represent the results of analytical and simulation models, respectively. It clearly shows that two models have the approximate results, although there exists some coupling deviation between them. The deviation arises as a result of two additional coupling cases considered in the simulation model as coupling crosstalk. It should be noted that the coupling crosstalk has negligible effect on the x-polarized input, as shown in Fig. 5A(b). However, it gives rise to power loss of about 2% into SMF1 for the y-polarized mode propagating in SMF1, as shown in Fig. 5A(c). We present two kinds of results for power output from the ports 2 and 3, under the input condition of polarization orientation from 0° (x-polarization) to 90° (y-polarization), and return to 180° (x-polarization), as shown in Fig. 5C(b). The minor deviation between two kinds of results is cased by the coupling crosstalk, especially for the y-polarization input.

Remarkably, the coupling crosstalk can be greatly reduced when increasing the distance between the ring core and the sing-mode cores. The detailed analyses are given in the supplementary. To obtain the results with high accuracy, hereinafter we adopt the simulation model to analyze the coupling property of these three-core coupling structures. We show in Fig. 5C(c) the working bandwidth dependence of this polarization splitter on the core distances. In the case of $d = 6$ μm, the central wavelength shifts slightly and the coupling efficiency also decreases to some extent due to the close distance and resultant large coupling crosstalk. By increasing the core distances, despite that the coupling bandwidth of polarization splitting reduces, the coupling crosstalk can be effectively restrained. Considering the fabrication or tapering errors for the core radius, we study the effect of these errors on the central wavelength shift due to the dependence of the resonance condition on waveguide dispersion. As shown in Fig. 5C(d), if the radius of ring core has design errors of ±4 nm, there will be a wavelength shift of about 2 nm.

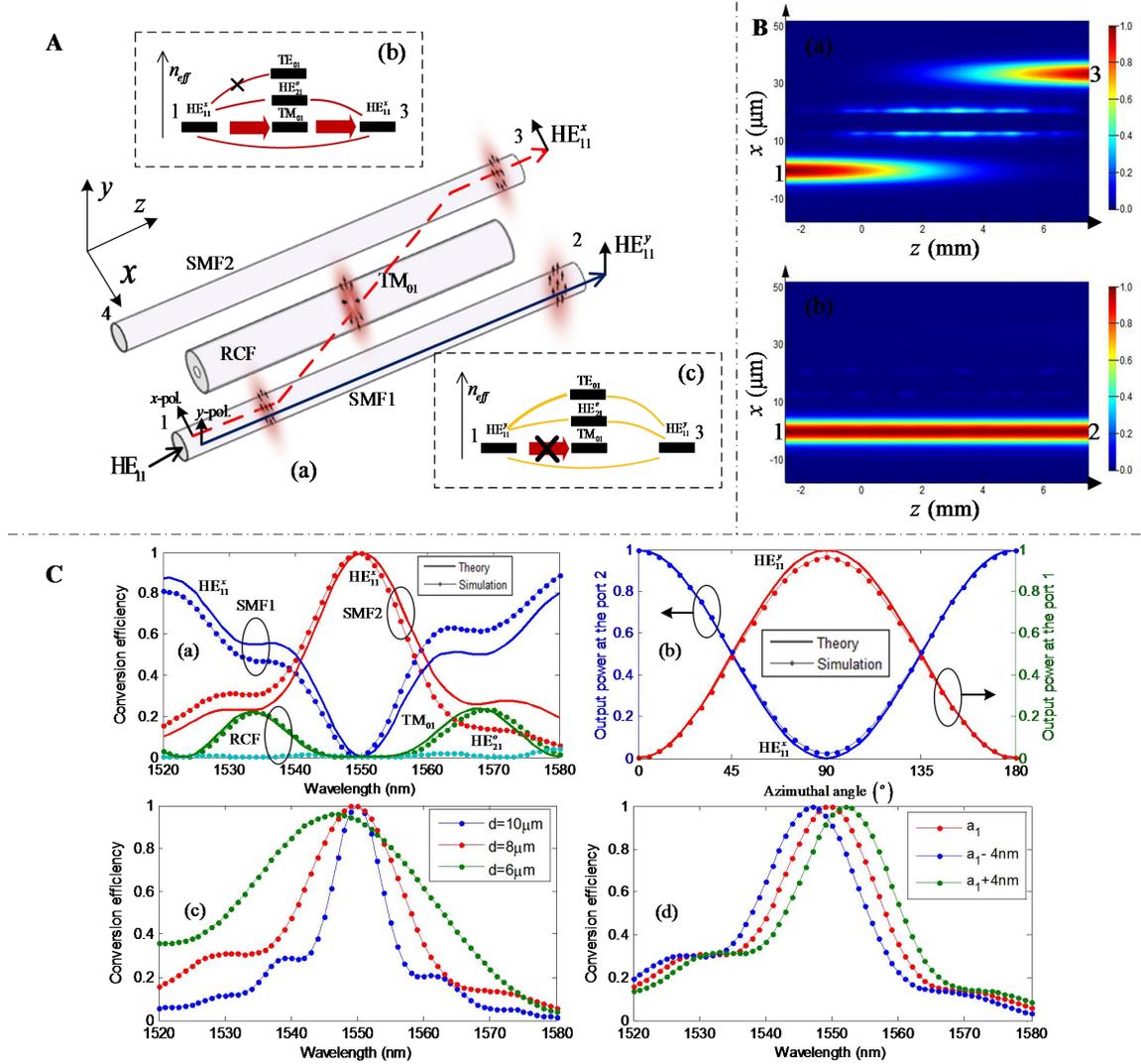

Fig. 5 **A** (a) Sketch of all-fiber polarization splitting structure and coupling paths for different polarization input, (b) power transfer among different modes for x-polarization input from port1 to port3, (c) coupling crosstalk for y-polarization input from port1 to port3. **B** Simulated power evolution map at the cross section of x-z for (a) x-polarization input, and (b) y-polarization input. **C** (a) Two types of calculated results on conversion efficiency for x-polarization splitting, (b) power output from ports 2 (left y-axis) and 3 (right y-axis) for polarization orientation from 0° to 180°, (c) simulated conversion efficiency for x-polarization splitting under different core distances, (d) simulated conversion efficiency for x-polarization splitting under two fabrication or tapering errors for ring core radius relative to $a_1 = 2794$ nm.

**All-fiber polarization rotator**

The sketch of three core layout for polarization rotating structure is shown in Fig. 6A(a), of which the section view corresponds to Fig. 4B(b), and the coupling path and polarization evolution are indicated as well. It should be stressed that all the segregated $TM_{01}$, $TE_{01}$, and $HE_{21}$ modes can serve as the middle matched mode in the ring core, and here we just utilize the $TM_{01}$ mode. For $TM_{01}$ or $TE_{01}$ modes as the middle matched mode, from port1 to port3, the rotating polarization is correspondingly from x-polarization to y-polarization, and from y-polarization to x-polarization, respectively. However, for HE$_{21}$ mode as the middle matched mode, the rotating polarization can work for both two orthogonal polarization inputs, because for different polarization input, the excited middle HE$_{21}$ mode can spontaneously correspond to odd and even HE$_{21}$ modes due to nearly the same effective indices between them. We simulate this polarization rotating structure and present the power propagation map in Figs. 6**B**(a) and 6**B**(b). Fig. 6**B**(a) shows the map at the cross section of x-z for x-polarization input from port1, and Fig. 6**B**(b) shows that at the cross section of y-z for y-polarization output from port3.

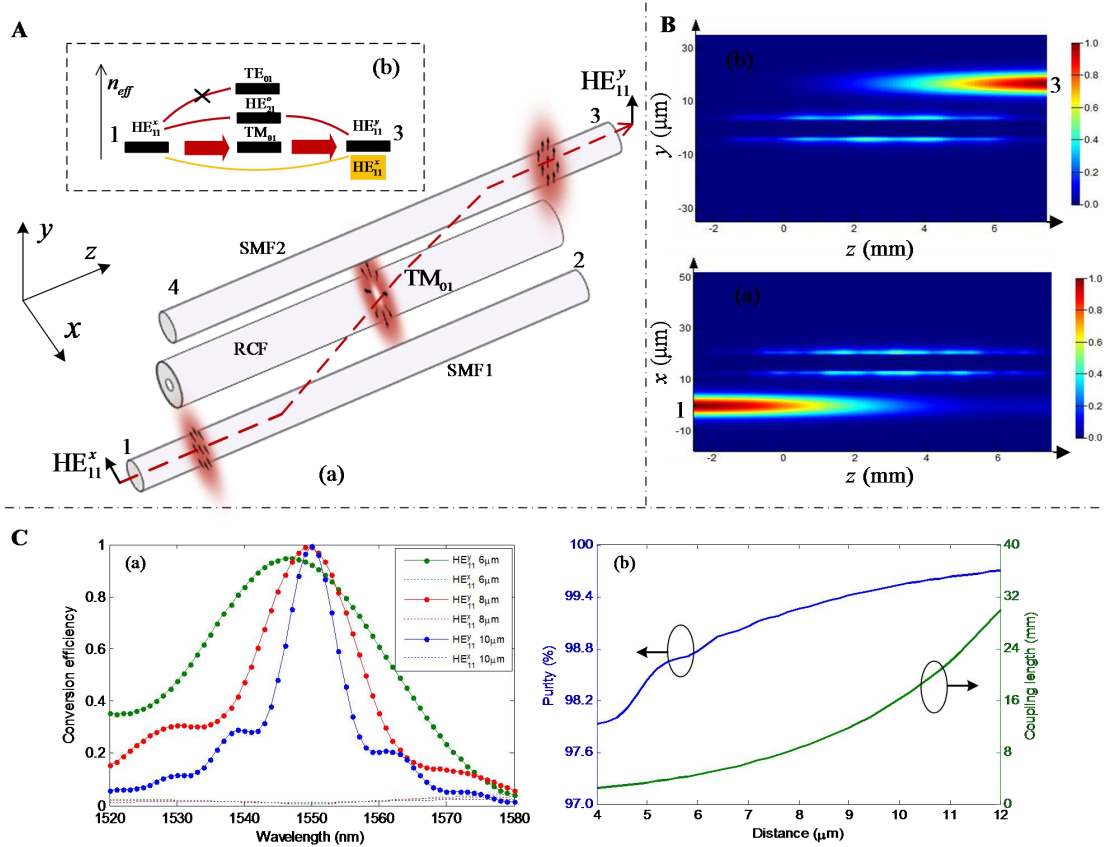

Fig. 6 **A** (a) Sketch of all-fiber polarization rotating structure and coupling paths for x-polarization input, (b) power transfer and coupling crosstalk among different modes for x-polarization input from port1 and y-polarization output from port3. **B** Simulated power propagation map (a) at the cross section of x-z for x-polarization input, and (b) that of y-z for y-polarization output. **C** (a) Simulated results of conversion efficiency for polarization-rotated HE$_{11}$ mode as well as coupling crosstalk under different core distances of 6 μm, 8 μm, and 10 μm, respectively, (b) the purity of polarization rotation (left y-axis), and required coupling lengths (right y-axis) versus different core distances from 4 μm to 12 μm.

We investigate the direct coupling of the x-polarized HE$_{11}$ from port1 to port3 as coupling crosstalk that affects the purity of the polarization rotation. As denoted in Fig. 6**A**(b), the yellow thin line represents the path of coupling crosstalk. It cannot be ignored because the quantity of coupling crosstalk possibly become remarkable due to the closer distance between two sing-mode cores, compared with the three-core layout of polarization splitting structure. We get the simulated results in Fig. 6**C**(a) under three different core distances (6μm, 8μm, and 10μm) between the middle ring core and two sing-mode cores. One can see that the coupling crosstalk isnot obvious over a wide wavelength range so that we can consider that the purity of polarization rotation is high at the resonant wavelength. We present this purity as a function of the core distances as indicated in the left y-axis in Fig. 6**C**(b). It can be quantified by $\eta = A_y^2 / (A_x^2 + A_y^2) \times 100\%$, where $A_x$ and $A_y$ stand for the magnitude of electric field of the coupled x-polarized and y-polarized $HE_{11}$ modes, respectively. It shows that the purity of polarization rotation reaches above 98 %, and can be further enhanced with increased the core distance. The required coupling length is also correspondingly given in the right y-axis in Fig. 6**C**(b). It indicates that longer coupling length is needed to compensate the weak coupling strength when increasing the core distances. In this case, it can be also predicted that the working bandwidth will be reduced accordingly.

**All-fiber polarization exchanger**

Finally, we study the all-fiber-based polarization exchange between two different transmission channels. In this design, the cores layout is the same as the polarization rotating structure, and its section view also corresponds to Fig. 4**B**(b). Although the polarization rotator discussed above can be used to exchange polarization between two channels when inputting x-polarized $HE_{11}$ mode at the port1 and meanwhile y-polarized $HE_{11}$ mode at the port4, these $TM_{01}$ and $TE_{01}$ middle matched rotators can only work for this specified orthogonal polarization pair, but does not work for the reverse polarization pair input, i.e., input y-polarization at the port1 and meanwhile x-polarization at the port4 in Fig. 6**A**(a). Therefore, to achieve the reciprocal polarization exchange for two different channels, we adopt the design of $HE_{21}$ middle matched mode in the ring core, as shown in Fig. 7**A**(a). In this case, the inner core radius of RCF needs to be fabricated or tapered to the marked value $a'$ in Fig. 3**A** to make the coupling resonance between $HE_{11}$ mode in two SMFs and $HE_{21}$ in RCF. The channels 1 and 2 correspond to y-polarization input at the port1 and meanwhile x-polarization input at the port4. This orthogonal polarization pair excites even $HE_{21}$ mode in the ring core, and couples to its contrary polarization, i.e. exchanging polarization between channels 1 and 2. The coupling crosstalk may be induced by $TE_{01}$ mode in the ring core as shown in Figs. 7**A**(b) and (c), but it is negligible according to the analyses of above polarization rotator. For the reverse polarization pair input denoted as channels 3 and 4, i.e. y-polarization input at the port1 and meanwhile x-polarization at the port4, it spontaneously excites odd $HE_{21}$ mode in the ring core, and also exchange the polarization directions between these two channels. Analogously, the negligible coupling crosstalk is induced by the $TM_{01}$ mode , which is not indicated in this figure.

We simulate the polarization exchanging structure, and verify the input and output modal power distribution and polarization evolution in Fig. 7**B**. Figs. 7**B**(a) and (b) correspond to the polarization exchanging between channels 1 and 2, whereas Figs. 7**B**(c) and (d) to channels 3 and 4. The polarization rotation for two orthogonal polarization inputs has a uniform azimuthal orientation because of the same phase difference through the coupling region, based on Eqs. (4)-(6). Actually, apart from the function of polarization exchange between two channels, if it is used as the polarization rotating device, this $HE_{21}$ middle matched coupling design rotates polarization direction with an orientation of $\pi/2$ for arbitrary polarization direction input, compared with the $TM_{01}$ or $TE_{01}$ middle matched polarization rotator that only works for a single polarization direction as discussed above.

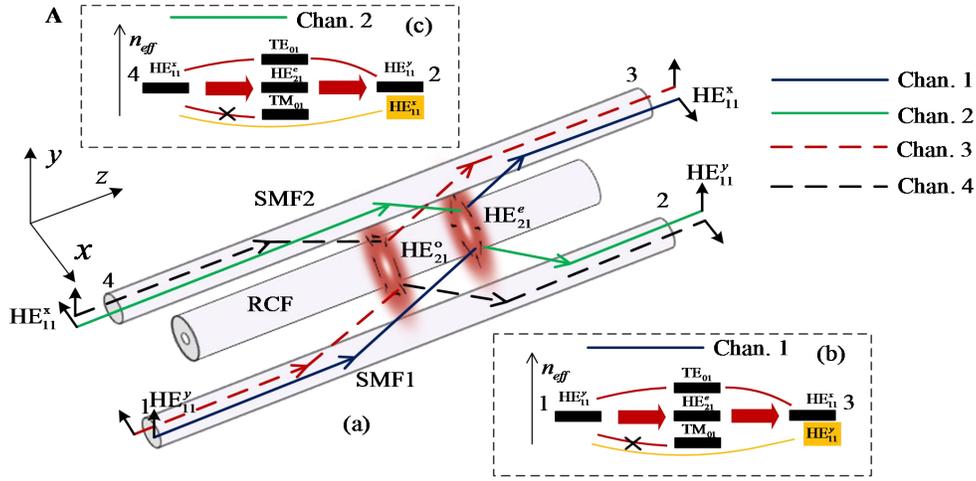

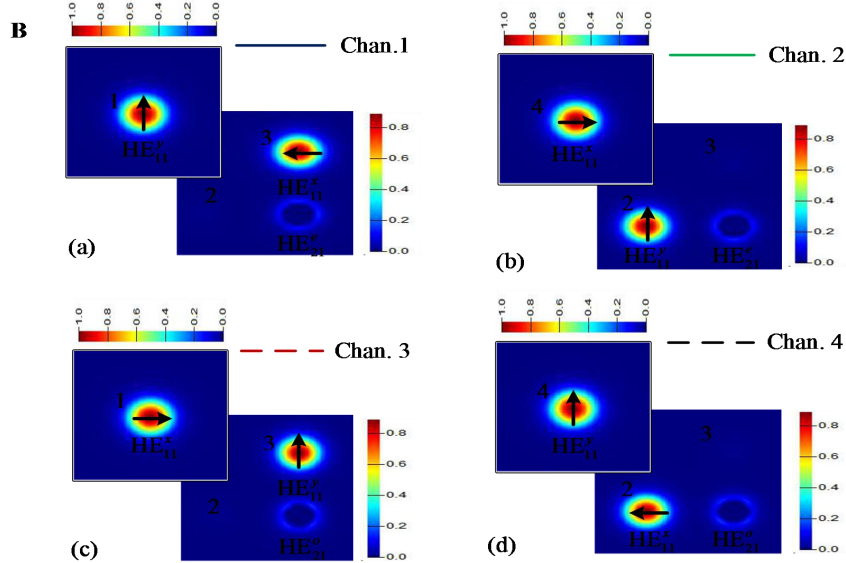

Fig. 7 **A** (a) Sketch of all-fiber polarization exchanging structure and coupling paths for different polarization channels, (b) power transfer and coupling crosstalk among different modes for y-polarization input from port1 and x-polarization output from port3 along the channel1, (c) power transfer and coupling crosstalk among different modes for x-polarization input from port4 and y-polarization output from port2 along the channel2. **B** Simulated input and output modal power distribution for 4 different channels, (a) y-polarized $HE_{11}$ input from port1 and x-polarized $HE_{11}$ output from part3 for channel1, (b) x-polarization input from port4 and y-polarization output from part2 for channel 2, (c) x-polarization input from port1 and y-polarization output from part3 for channel3, (d) y-polarization input from port4 and x-polarization output from part3 for channel4. The arrows represent the polarization direction of coupled $HE_{11}$ modes.

**Discussion**

Here, the first vector mode group ($TM_{01}$, $TE_{01}$, and $HE_{21}$ vector modes) is exploited as the middle matched mode of ring core in the three-core fiber couplers to achieve optical polarization split, rotation, and exchange. For higher-order mode groups, such as the group of $HE_{31}$ and $EH_{11}$ modes, or $HE_{41}$ and $EH_{21}$ modes, etc., they cannot be used to split polarization due to nearly the same effective indices between even and odd vector modes, nevertheless can be utilized to rotate and exchange polarization. However, on one hand, it may induce larger coupling crosstalk, compared with the first vector modes group because of the reduced effective refractive index differences between two vector mode components within the higher-order mode groups. On the other hand, the spatial orientation period of polarization variation becomes smaller and smaller when increasing the radial mode order so that the purity of polarization rotation may be sharply reduced. For the polarization-selective mode splitting, apart from RCF, any high-contrast index fibers that can lift the modal degeneracy can be utilized, as long as the effective refractive indices of the odd and even vector mode components can be segregated large enough. For instance, the high-contrast index elliptical core fiber can also be adopted as the middle fiber to split polarization, because this kind of elliptical core can easily introduce modal birefringence between two orthogonal vector components, and thus filter one specific polarization coupling. These polarization-selective coupling design might be extended to all-fiber-based multiplexer/demultiplexer of higher-order LP modes with two orthogonal polarization states in FMFs instead of using the multi-input and multi-output (MIMO) technology,[11] as provided in the supplementary.

The most remarkable application of the proposed three-core fiber directional couplers is its ability to arbitrarily get polarization-desired coupling that is pivotal to all-fiber-based polarization management in fiber PDM and fiber laser systems, such as polarization split and isolation, polarization rotation and exchange for specific polarization division channels, as well as polarization multiplexing/demultiplexing. Perhaps, they can also be applied to fiber sensing based on the measurement of polarization variation induced by the physical parameter change.

**Conclusions**

In summary, we present an all-fiber design of three-core coupling structure to achieve flexible polarization management for fiber-guided modes. We take advantages of the spatially vectorial distribution of fiber-guided $TM_{01}$, $TE_{01}$, and $HE_{21}$ modes in a high-contrast index ring core that can support these segregated cylindrical vector modes. By embedding this ring core between conventional fiber coupler, we can create the polarization-selective coupling between two SMFs or two channels. We design three types of polarization-manageable couplers, i.e. polarization splitter, rotator, and exchanger. We investigate the coupling properties of these coupling structures in detail based on the coupled-mode theory and numerical simulations. The obtained results manifest favorable operation performance of comprehensively managing the fiber modal polarization states. Compared to previous management of power, wavelengths and modal forms, these kinds of all-fiber-based couplers can realize an additional manageable degree of freedom for modal polarization. They may provide an opportunity for potential all-fiber-based optical polarization split, isolation, exchange, and multiplexing/demultiplexing in fiber communications and sensing.

**Acknowledgments**

This work was supported by the National Program for Support of Top-Notch Young Professionals, National Basic Research Program of China (973 Program) under grant 2014CB340004, the National Natural Science Foundation of China (NSFC) under grants 11574001, 11274131 and 61222502, and the Program for New Century Excellent Talents in University (NCET-11-0182).